\documentstyle[prl,aps,epsfig,multicol]{revtex}
\begin{document}
\def\be{\begin{equation}}
\def\ee{\end{equation}}
\def\bearr{\begin{eqnarray}}
\def\eearr{\end{eqnarray}}
\def\la{\langle}
\def\ra{\rangle}
\def\l{\left}
\def\r{\right}
\draft
\title{ Quasi-one dimensional degenerate Bose gases}
\author{Tarun Kanti Ghosh}
\address
{ The Abdus Salam International Centre for Theoretical Physics, 
34100 Trieste, Italy.}

\date{\today}
\maketitle

\begin{abstract}
Motivated by the MIT experiment [Gorlitz {\em et al.}, Phys. Rev. Lett. {\bf 87}, 
130402 (2001)], we analytically study the effect of density and phase fluctuations 
on various observables in a quasi one-dimensional degenerate Bose gases.
Quantizing the Gross-Pitaevskii Hamiltonian and diagonalize it in terms of 
the normal modes associated with the density and phase fluctuations of a quasi-one 
dimensional Bose gas.
We calculate dynamic structure factor $S(q,\omega )$ from low-energy condensate density
fluctuations and find that there are multiple peaks in 
$S(q,\omega )$ for a given momentum $q$ due to the discrete
spectrum.  These multiple peaks can be resolved by a two-photon Bragg pulse with a long
duration which transfer the momentum to the system.  We calculate the momentum
transferred $ P_z(t)$ by using the phase-density representation of the Bose order parameter.
We also calculate the single-particle 
density matrix, phase coherence length, and momentum distribution
by taking care of the phase fluctuations upto fourth-order term as well as 
the density fluctuations.
Our studies on coherence properties shows that 1D Bose gases of MIT experiment
do not form a true condensate, but it can be obtained by a moderate changes of the 
current experimental parameters.

\end{abstract}


\begin{multicols}{2}[]

\section{Introduction}
Recently, there have been a great interest to study theoretically \cite{ket,ho}
as well as experimentally \cite{gor} on the low-dimensional Bose 
gases. Among the 
various interesting topics related to the trapped Bose-Einstein condensation (BEC), 
the study of  structure factors and  coherence properties of quantum Bose gas 
has attracted major interest. The dynamic structure factor plays an important role
which helps us to understand the discrete modes due to the finite size system \cite{tozzo}.
Unfortunately, the dynamic structure factor calculated from the local density approximation
do not show up the discrete nature of the modes and this is valid under
certain conditions \cite{dal,ket1}. 
The dynamic structure factor can be analyzed through the behavior of the momentum
transferred to a trapped Bose gases by two-photon Bragg pulse with long duration \cite{stein} which reflects 
the underlying discrete spectrum \cite{tozzo}.
On the other hand, coherence has importance for many applications, like matter wave
interferometry, guided atomic beams and atom lasers. 
The realization of the quasi-one dimensional Bose systems at MIT \cite{gor} raises the question
of the effects of quantum and thermal fluctuations. 
The discrete nature of the excitations implies that BEC will not be 
destroyed immediately as interactions between bosons are turned on \cite{ket,ho}. 
In low dimensions, long-wave length density and phase fluctuations 
lead to a new intermediate state between a condensate and 
non-condensate system. The system forms a {\em quasi-condensate}, in which the phase 
of the order parameter is only coherent over a finite distance less than the 
system size, whereas the phase is coherent over a distance of the order 
of the size of the system in a true condensate.
Quasi-condensate of a phase fluctuating BECs have been the subject of theoretical
studies \cite{pet1,pet2,kag2,pet3}, including the development of a 
improved many-body $T$-matrix  theory valid in all dimensions and at low temperatures 
\cite{stoof1,stoof2}, and the extension of the Bogoliubov theory to low-dimensional 
degenerate Bose gases \cite{mora}. Also, the phase fluctuating nature of highly elongated BEC 
was experimentally demonstrated in Ref. \cite{exp1,exp2,exp3}.

In this paper, we investigate in details on the properties of the effect of the low-energy modes,
particularly, dynamic structure factor, and also the momentum transferred by a two-photon Bragg pulse since 
there have not been any study in the literature on these properties. Moreover,
in the theoretical studies on degenerate Bose gases have neglected the contribution from the 
density fluctuations and have concentrated only upto the quadratic phase fluctuations. 
The density fluctuations and higher order phase fluctuations are important to study coherence
properties for a wide range of the system parameters. 
In this work, we calculate one-body density matrix, phase correlation length and momentum distribution 
by taking care of the density fluctuations and fourth-order phase fluctuations. 

This paper is organised as follows. In section II, using the phase-density representation of the
Bose order parameter, we write down an effective hydrodynamic Hamiltonian 
and diagonalize it in terms of the normal modes of the density and phase fluctuations.
First, we rederive the known low-energy modes at zero temperature. 
Then we incorporate the temperature into theory through the temperature dependent chemical potential.
In section III, we calculate static and dynamic structure factors.
We also present an analytic expression for the time-evolution of momentum transfer to the system
due to the Bragg pulse. 
In section IV, we consider the effect of the density fluctuations and the fourth-order 
term in the phase fluctuations and calculate the phase coherence function, 
the phase correlation length and momentum distribution. We also discuss the possibility to form a 
true condensate by changing the system parameters. In section V, we give a brief summary 
of our work.

\section{quantization of 1D Bose gas}
We consider atomic Bose gases confined in a quasi-one dimensional
harmonic trap as described in MIT lab \cite{gor}. For all the Figs. in this paper, 
we will be using the parameters used at the MIT experiment on
quasi-one dimensional BEC by Gorlitz {\em et al.} \cite{gor}. In particular, they have
used $^{23}$ Na in the trap with axial trapping frequency $ \omega_z = 2 \pi \times 3.5 $ Hz, 
($ a_z = 1.12 \times 10^{-5}$ m)
and radial trapping frequency $ \omega_0 = 2 \pi \times 360 $ Hz, ($a_0 = 1.10 \times 10^{-6}$ m).
The three-dimensional $s$-wave scattering length is $ a = 2.75 \times 10^{-9} $ m. The number of
condensate atoms varies from $ N_{c0} \sim 10^4 $ to $ N_{c0} \sim 10^5 $ which will be specified in the Figs.
The chemical potential at zero temperature is 
$ \mu_0 = \hbar \omega_z (1.06 \frac{aa_z}{a_0^2} N_{c0})^{2/3} \sim 40 \hbar \omega_z $ for $ N_{c0} \sim 10^4 $.
The parameters satisfy the condition for the system to be quasi-one dimensional: $ \hbar \omega_0 >> \mu_0 >> 
\hbar \omega_z $. 

The grand-canonical Hamiltonian of the Bose system is
\bearr
\hat H^{\prime} & = & H - \mu N \nonumber 
\\ & = & \int dz \hat \psi^{\dag} (z,t)( - \frac{\hbar^2}{2m} \frac{\partial^2}{\partial z^2} +
+\frac{1}{2} m \omega_z^2 z^2 - \mu )  \hat \psi (z,t) \nonumber
\\ & + & \frac{g_{1}}{2} \int dz \hat \psi^{\dag} (z,t) \hat \psi^{\dag} (z,t) \hat \psi (z,t) \hat \psi 
(z,t).
\eearr 
The equation of motion for the bosonic field operator 
\bearr
i \hbar \frac{ \partial \hat \psi (z,t)}{\partial t} & = &  
[ \hat \psi (z,t), \hat H^{\prime}] \nonumber 
\\ & = & [- \frac{\hbar^2}{2m} \frac{\partial^2}{\partial z^2}
+\frac{1}{2} m \omega_z^2 z^2 - \mu ] \hat \psi (z,t) \nonumber
\\ & + & g_{1} \hat \psi^{\dag}(z,t) \hat \psi (z,t) \hat \psi (z,t),
\eearr
where the effective coupling strength is $ g_1 = 2 \hbar \omega_0 a $.
The Bose field operator can be decomposed in the as usual way,
\be
\hat \psi (z,t) = \Phi (z,t) +  \tilde \psi (z,t),
\ee
then the equation of motion for the condensate wave function is
\bearr \label{gpva}
i \hbar \frac{ \partial \Phi (z,t)}{\partial t} & = & 
[- \frac{\hbar^2}{2m} \frac{\partial^2}{\partial z^2}
+\frac{1}{2} m \omega_z^2 z^2 - \mu ] \Phi (z,t) \nonumber
\\ & + & g_{1} |\Phi (z,t)|^2 \Phi (z,t) + 2 g_{1d} \Phi (z,t) \tilde n(z,t) \nonumber
\\ & + & g_{1} \Phi^* (z,t) \tilde m(z,t). 
\eearr
Here, $ \tilde n(z,t) = \la \tilde {\psi}^{\dag}(z,t) \tilde {\psi}(z,t) \ra $ and 
$ \tilde m(z,t)= \la \tilde {\psi} (z,t) \tilde {\psi} (z,t) \ra $ are the normal and 
anomalous non-condensed particle densities, respectively.
When temperature is very low, one can neglect $ \tilde n(z,t) $ and $ \tilde m(z,t) $, then
the equation reduces to the usual Gross-Pitaevskii equation:
\bearr \label{hamil}
H^{\prime} & = & \int dz \Phi^* (z,t) [ - \frac{\hbar^2}{2m} \frac{\partial^2}{\partial z^2}
+\frac{1}{2} m \omega_z^2 z^2 - \mu ] \Phi (z,t) \nonumber
\\ & + & \frac{g_{1}}{2} \int dz |\Phi (z,t)|^4.
\eearr  

The condensate density is
\be \label{density}
n_0 (z) = \frac{1}{g_1} [\mu_0 - \frac{1}{2} m\omega_z^2 z^2 ].
\ee

We use the quantized hydrodynamic theory developed by Wu and Griffin \cite{wu} at $T=0$.
The Bose order parameter can be written as $ \Phi(z,t) = \sqrt{n(z,t)} e^{i\phi(z,t)}$.
The fluctuations in the density and phase about their equilibrium are 
$ \hat n(z,t) = n_0(z) + \delta \hat n(z,t) $ and $ \hat \phi(z,t) = \phi_0 (z) + \delta \hat \phi(z,t) $. 
By keeping up to second-order in the fluctuations, we obtain the effective hydrodynamic Hamiltonian at
zero temperature,
\be
H = H_0 + \frac{1}{2} \int dz \l [m n_0(z) \delta \hat v_z^2(z,t) + g_1 \delta \hat n^2(z,t) \r],
\ee
where $ \delta v_z (z,t) = \frac{\hbar}{m} \frac{d}{dz} \delta \phi(z,t) $ is the superfluid velocity. 
This Hamiltonian is quadratic in the density and the phase fluctuation operators, then
we can diagonalize it using the canonical transformations:
\be
\delta \hat n(z,t) = \sum_{j} \l [ a_j \psi_j(z) e^{-i \omega_j t} \hat \alpha_j + H. c. \r],
\ee
and
\be
\delta \hat \phi(z,t) = \sum_{j} \l [ b_j \psi_j(z) e^{-i \omega_j t } \hat \alpha_j + H.c. \r],
\ee
The operators $ \hat \alpha_j $ and $ \hat \alpha_j^{\dag} $ destroy and 
create excitations with energy $ \hbar \omega_j$
and satisfy the commutation relations $[ \hat \alpha_j, \hat \alpha_{j^{\prime}}^{\dag}] = 
\delta_{j,j^{\prime}} $, 
$ [ \hat \alpha_j, \hat \alpha_j] = 0 $ and $ [ \hat \alpha_j^{\dag}, \hat \alpha_j^{\dag}] = 0 $.
Also, $ [\delta \hat n(z,t), \delta \hat \phi (z^{\prime},t)] =  i \delta (z - z^{\prime}) $.
The constant terms are $ a_j = i \sqrt{\frac{\hbar \omega_j}{2g_1}}$ and 
$b_j = \sqrt{\frac{2g_1}{\hbar \omega_j}}$.
The equations for the density and phase fluctuations can be obtained by using the Heisenberg equation
of motion:
\be
\frac{ \partial^2 \delta \hat n(z,t)}{\partial t^2} = \frac{\partial}{\partial z} 
\l [\frac{g_1}{m} n_0(z) \frac{\partial}{\partial z} \delta \hat n(z,t) \r],
\ee
and 
\be
\frac {\partial^2 \delta \hat \phi (z,t)}{\partial t^2} = \frac{\partial}{\partial z}
\l [\frac{g_1}{m} n_0(z) \frac{\partial}{\partial z} \delta \hat \phi (z,t) \r].
\ee

The equation for the eigenfunctions $ \psi_j (z) $ is
\be \label{gen1}
\nonumber \frac{g_1}{m}\frac{\partial}{\partial z} \l [ n_0(z) \frac{\partial}{\partial z} \psi_j (z) \r ] 
+ [\omega_j^2 ]\psi_j (z) = 0.
\ee
Therefore, Eq.(\ref{gen1}) becomes Legendre equation
with the eigenfrequencies given by $ \omega_j = \omega_z \sqrt{j(j+1)/2} $ and the
corresponding normalized eigenfunctions are $ \psi_j (z) = \sqrt{\frac{2j+1}{2 Z_0}} P_j(z/Z_0) $,
where $ P_j(z/Z_0) $ is the Legendre polynomial in $z$ and $Z_0 $ is the Thomas-Fermi half length
\cite{ho}.
The breathing mode corresponds to the $j=2 $ which oscillates with the frequency
$\omega_2 = \sqrt{3} \omega_z$ and it has also been verified experimentally \cite{moritz}. 
The chemical potential depends on the number of particles in the condensate and the 
the number of particles in the condensate strongly depends on the temperature, 
$ N_c(T) = N_{c0}(1-T/T_c) $,
where $T_c $ is the critical temperature of an ideal Bose gases \cite{ket}. 
The simplest way of including the non-condensate is to use the temperature dependent 
condensate number $N_c(T)$ in the chemical potential and therefore in the Thomas-Fermi half-length. 
This approximation is similar to the local density approximation.
Therefore, the  chemical potential at finite temperature can be written as
\be
\mu = \hbar \omega_z [1.06 \frac{aa_z}{a_0^2} N_{c0}(1-\frac{T}{T_c})]^{2/3},
\ee
and then the Thomas-Fermi half-length at finite temperature $T$ can be obtain from
$  m \omega_z^2 Z_T^2 = 2 \mu $. This temperature dependent Thomas-Fermi length, $Z_T$,
will be useful to study the coherence properties.
The condensate normal mode frequencies at finite $T$ are 
the same as at $T=0$ in the Thomas-Fermi limit, since the frequencies at $T = 0$ do not 
depend on the value of $N_c $ in this limit, but the eigenfunctions changes through the 
TF half-length. We note that the amplitude of the quasi-particle increases as temperature 
increases.

\section{structure factors and Bragg spectroscopy}
{\em Dynamic Structure Factor}:
Consider a low-intensity off-resonant inelastic
light scatters with momentum transfer $\hbar q $ and energy transfer $\hbar \omega $
to the target (BEC). If the external light couples weakly to the number density
of the target, the differential cross section is proportional to the dynamic structure factor
$ S(q,\omega) $, which is obtained from the Fourier transform of the time-dependent density-density
correlation functions,
\be
S(q,\omega)  = \int dt \int dz e^{i(\omega t - qz)} < \delta \hat n(z,t) \delta \hat n(0,0)>.
\ee
It is the density fluctuation spectrum that can be measured in the two-photon Bragg 
spectroscopy. 
The dynamic structure factor of this system can be written as,
\bearr
S(q,\omega) & = & \sum_{j=1} \frac{\hbar \omega_j}{2 g_1} |\psi_j(q)|^2 \nonumber \\ 
& \times & [ f_j \delta(\omega+\omega_j) +(1+ f_j ) \delta(\omega - \omega_j)],
\eearr
where $ f_j = [exp{(\beta \hbar \omega_j)} - 1]^{-1} $ is the thermal Bose-Einstein
function and $ \psi_j(q) = \int dz e^{-iqz} \psi_j(z/Z_T) $ is the spatial Fourier transform of 
$ \psi_j(z) $. 
It can be easily shown that 
\be
|\psi_j(\tilde q)|^2 =   \pi (2j+1) Z_T \frac{[J_{j+\frac{1}{2}}(\tilde q)]^2}{ \tilde q},
\ee
where the dimensionless variable is $ \tilde q = q Z_T$.
It is convenient to rewrite the dynamic structure factor as
$$ 
\nonumber{S( \tilde q,\omega) =  \sum_{j} S_j(\tilde q) [f_j \delta(\omega+\omega_j) 
+ (1+ f_j ) \delta(\omega - \omega_j)]},
$$
where
\be \label{sjq}
S_j(\tilde q) = \frac{\pi}{4 \sqrt{2}} \frac{ a_0^2}{a a_z} \frac{Z_T}{a_z} (2j+1) \sqrt{j(j+1)}
\frac{[J_{j+\frac{1}{2}}(\tilde q)]^2}{ \tilde q}.
\ee
These functions determine the weight of the light-scattering cross-section in $S(q,\omega)$
of the corresponding collective modes of energy $\omega_j$.
In Fig.1 we have plotted $ S_j(\tilde q) $ which as a function
of the dimensionless wave vector $ \tilde q = qZ_0 $ for the excitations $ j = 1,2,3,4,$ and $ 5 $.
Fig.1 shows how many modes significantly contribute to $S(q,\omega)$. Note that $j=0$ mode do not contribute
to $S(q,\omega)$.
It is clear from the Fig.1 that the strongest weights for these collective modes appear for
$ qZ_0 \geq 2 $. As an example, for the parameters used at MIT experiment \cite{gor} 
which gives $ Z_0 \sim 9 a_z $, this means that the momentum transfer $q$ in a light scattering 
experiments should be $ q \geq 0.22 a_z^{-1} $ for $ N_{c0} = 10^4 $
in order to pick up the strong spectral weight from the low-energy collective modes.
We also notice that the number of modes that contribute to the $S(q,\omega)$ increases with 
the chemical potential.

\begin{figure}[h]
\epsfxsize 9cm
\centerline {\epsfbox{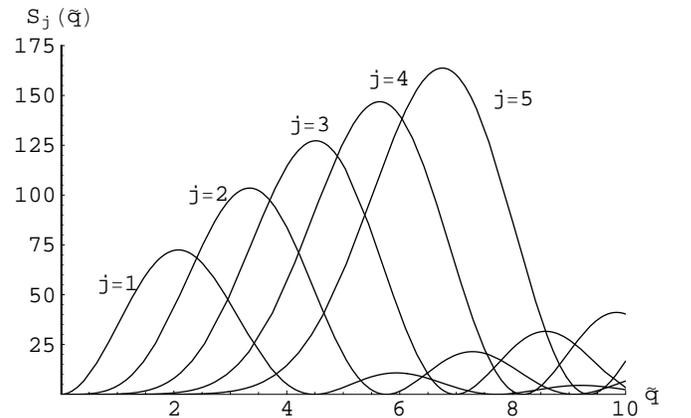}}
\vspace{0.2 cm}
\caption{Plots of $S_j( \tilde q) $ as a function of the dimensionless wave
vector $\tilde q $.}
\end{figure}

For $ \tilde q \rightarrow 0$, the leading term arises from the Kohn or dipole-sloshing
mode $ j =1 $, with $S_1 \sim \tilde q^2$; the next contributions arise from the terms
with $j = 2, 3 $. 
In Fig.2 we plot the dynamic structure factor $S( \tilde q,\omega)$
for the momentum transfer corresponds to $ \tilde q = qZ_0 = 4 $.
For finite-energy resolution we have replaced the delta function by the Lorentzian 
with a width of $ \Gamma = 0.06 \omega_z$.
\begin{figure}[h]
\epsfxsize 9cm
\centerline {\epsfbox{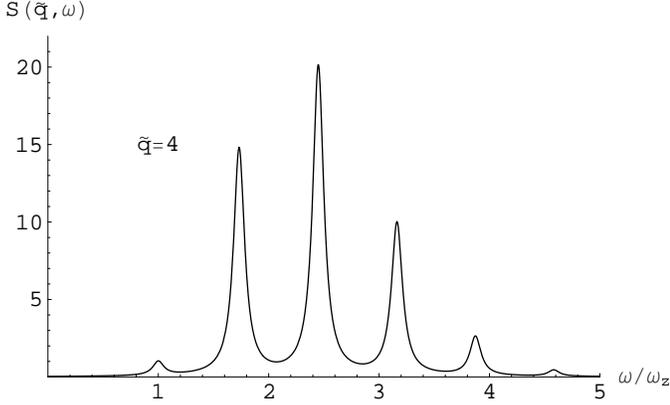}}
\vspace{0.2 cm}
\caption{Plot of the dynamic structure factor $ S(q,\omega)$ vs. $\omega$ at $T=0$ with 
momentum transfer $ \tilde q = 4 $.} 
\end{figure}

In Fig.3 we present the results for a higher momentum
transfer $ \tilde q = 10 $ using the same arbitrary units as in the Fig.2 ,
showing multiple peaks at higher energy transfer $\omega $. This is expected since for large $\tilde q$, the
wave-length is small compared to the system size and it excites the higher energy modes.

\begin{figure}[h]
\epsfxsize 9cm
\centerline {\epsfbox{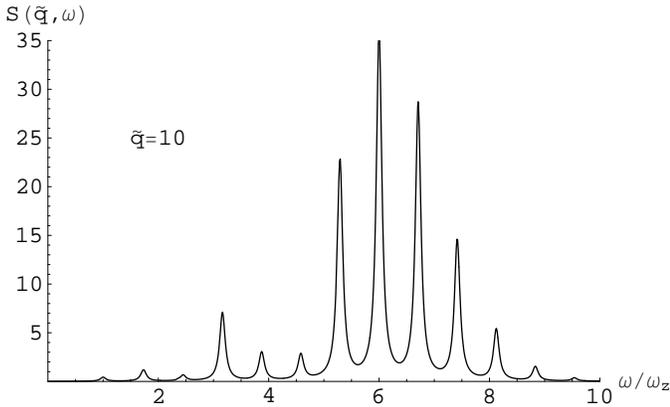}}
\vspace{0.2 cm}
\caption{Plot of the dynamic structure factor $ S(q,\omega)$ vs. $\omega$ at $T=0$ for
momentum transfer $ \tilde q = 10$.}
\end{figure}

As one can see from the Fig.2 and Fig.3, the dynamic structure factors 
has multiple peaks and the number of peaks increases if the given momentum transferred is high. 
This phenomena is due to the underlying discrete spectrum.
Note that the energy transfer $ \omega $ is much smaller than the gap between the chemical potential
$\mu$ and the transverse excited state ($\sim \omega_0$) and hence the radial modes can
not contribute to the dynamic structure factor. 
This behavior of the multiple peaks can be resolved in two-photon Bragg spectroscopy, as
shown by Steinhauer {\em et al.} \cite{stein}. 

{\em Bragg Spectroscopy}:
The observable in the Bragg scattering experiments is the momentum transferred to the
condensate. When the system is subjected to a time-dependent Bragg pulse, the additional
interaction term appears in the Hamiltonian which is given by \cite{blak},
\be
H_I(t) = \int dz \hat \psi ^{\dag} (z,t) [ V_B(t) cos(qz-\omega t)] \hat \psi (z,t).
\ee
Here, we supposed that the Bragg pulse is switched on at time $t=0$ and $q$ is also along the 
$z$-direction. The momentum transferred $P_z(t)$ by the Bragg pulse has been calculated based on the Bogoliubov 
transformation \cite{blak}. Here, we calculate $P_z(t)$ by using the phase-density representation
of the Bose order parameter.
After linearizing the Bose order parameter, the above term becomes,
\bearr
H_I(t) & = & \int dz n_0(z) V_B(t) cos(qz-\omega t) + \sum_j \hat \alpha_j e^{i \omega_j t} a_j \nonumber \\
& \times &  \int dz \psi_j(z)  V_B(t) cos(qz-\omega t) + H. c.
\eearr  
The second term in the above equation describe the scattering between the condensate
and the quasiparticles which occurs due to the energy and momentum transfer from the
optical potential. The time-dependent exponentials, $ e^{\pm i \omega_j t} $, which multiply the
quasiparticle operators in Eq. account for the free evolution due to the Hamiltonian $H^{\prime}$ [given 
in Eq. (\ref{hamil})], and so the Heisenberg equation of motion
\be
i \hbar \frac{\partial }{\partial t} (\hat \alpha_j e^{- i \omega_j t}) = 
[(\hat \alpha_j e^{- i \omega_j t}), H + H_I(t)]
\ee
becomes
\be
i \hbar \frac{\partial }{\partial t} \hat \alpha_j = [ \hat \alpha_j, H_I(t)].
\ee
After solving the equation,
\bearr
\hat \alpha_j (t) & = & \hat \alpha_j (0)  \nonumber \\ 
& + & \frac{i}{\hbar} \int_0^t dt^{\prime} V(t^{\prime}) e^{i \omega_j t^{\prime}}
\int dz a_j \psi_j(z) cos(qz -\omega t^{\prime}).
\eearr

The momentum transfer from the optical potential is
\bearr
P_z(t) & = & \sum_{j,q} \hbar q < \hat \alpha_j^{\dag} (t) \hat \alpha_j (t) >
 =  (\frac{V_B(t)}{2 \hbar})^2 \sum_j \hbar q S_j (\tilde q) \nonumber \\ 
& \times & F_j [(\omega_j - \omega),t] - F_j [(\omega_j + \omega),t],
\eearr
where $ F_j [(\omega_j \pm \omega),t] = (\frac{sin[(\omega_j \pm \omega)t/2]}{(\omega_j \pm \omega)/2})^2 $ and
$S_j (\tilde q) $ is given in Eq. (\ref{sjq}). For large $t$, $S(q,\omega) \sim P_z(t)$.
We plot $P_z(t)$ for various time in Fig.4 and Fig.5, and shows that the multipeak spectrum in $S(q, \omega)$ can be
resolved only when the duration of the Bragg pulse is $ t >> 2\pi/\omega_z $. 
When $ t < 2\pi/\omega_z $, $P_z(t)$ reflects the dynamic structure factor calculated from
the local density approximation.

\begin{figure}[h]
\epsfxsize 9cm
\centerline {\epsfbox{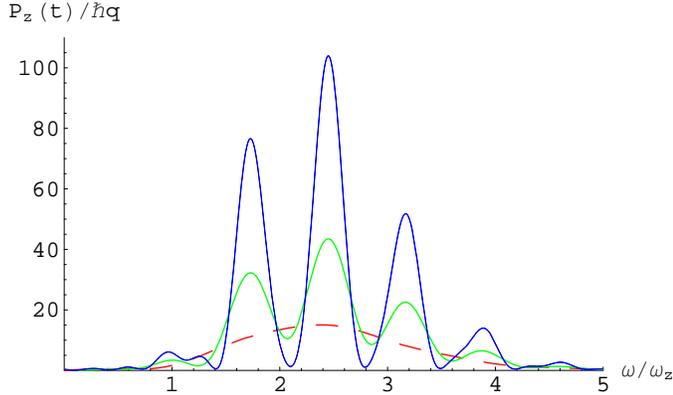}}
\vspace{0.2 cm}
\caption{Plots of the momentum transferred $ P_z(t)$ vs. $ \omega $ at $T=0$ when
momentum transfer is $ \tilde q = 4 $ for various time $ \omega_z t = 7.0 $ (dashed), $ \omega_z t = 12.0 $ (dotted) 
and $ \omega_z t = 19.0 $ (solid). Also, we have assumed $ V_B = 0.1 \hbar \omega_z$}.
\end{figure}

\begin{figure}[h]
\epsfxsize 9cm
\centerline {\epsfbox{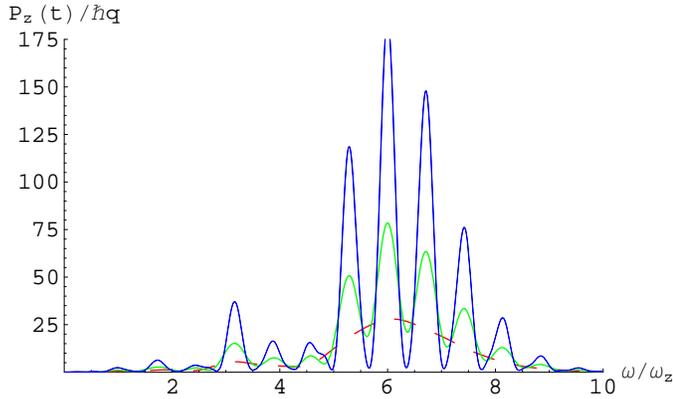}}
\vspace{0.2 cm}
\caption{Plots of the momentum transferred $ P_z(t)$ vs. $\omega $ at $T=0$ when
momentum transfer $ \tilde q = 10 $ for various time $ \omega_z t = 7.0$ (dashed), $ \omega_z t = 12.0 $ (dotted),
and $ \omega_z t = 19.0 $ (solid). Here, $ V_B = 0.1 \hbar \omega_z$.}
\end{figure}

{\em Static Structure Factor}: For completeness, we also calculate
the static structure factor which can be obtained from the Fourier transform of the equal-time
density-density correlation function
\bearr
S(q) & = & \int dz e^{iq(z-z^{\prime})} \la \delta \hat n(z) \delta \hat n(z^{\prime}) \ra \nonumber \\
& = & \sum_j S_j(\tilde q) coth(\frac{\beta \hbar \omega_j}{2})
\eearr
with the long-wavelength limit
$ S(\tilde q) \sim \tilde q^2 coth(\frac{\beta \hbar \omega_1}{2})$.

\section{Density matrix and momentum distribution}
The single-particle density matrix at equal-time
is defined as
\be \label{pd}
D_1(z,z^{\prime}) = \la \hat \psi^{\dag}(z,t) \hat \psi(z^{\prime},t) \ra.
\ee
The bosonic field operator $ \hat \psi (z,t) $ can be written in terms of the 
phase-density representation as $ \hat \psi (z,t) = e^{i \hat \theta(z)} \sqrt{ \hat n(z)} $.
Using the phase-density representation of the  bosonic field operators into Eq. (\ref{pd}), 
we get
\be
D_1(z,z^{\prime}) =  \la \sqrt{\hat n(z)} e^{- \hat \theta(z)} 
e^{i(\hat \theta(z^{\prime})} \sqrt{ \hat n(z^{\prime})} \ra.
\ee
Most of the theoretical studies (except \cite{mora}) have considered 
$ \delta \hat n(z) \sim 0 $ since $ \delta \hat n(z) \sim \frac{\hbar \omega_j}{g_1}$. 
This is true for large interaction strength, but for small interaction strength there is a finite
contribution to the density matrix. Therefore, we must include the contribution from 
the density fluctuations to calculate one-body density matrix for a wide range of 
the system parameters.
Expanding  square root of the density operator upto second order in the following way,
\bearr 
\sqrt{\hat n(z)} & = & \sqrt{ n_0(z) + \delta \hat n(z)} \nonumber \\
  & = & \sqrt{ n_0(z)} [ 1 + \frac{\delta \hat n(z)}{2 n_0(z)} 
- \frac{1}{8} (\frac{\delta \hat n(z)}{ n_0(z)})^2 + .... ].
\eearr 
Using the commutation relations for the phase and density operators and 
applying the Wick's theorem, then the single-particle
density matrix can be written as,
\bearr
D_1(z,z^{\prime}) & = & \sqrt{ n_0(z) n_0(z^{\prime})} 
\la (1- \frac{1}{8}[\frac{\delta \hat n(z)}{ n_0(z)} - 
\frac{\delta \hat n(z^{\prime} )}{ n_0(z^{\prime})} ]^2) \ra \nonumber \\
& \times & \la e^{-i[ \delta \hat \phi(z,t) - 
\delta \hat \phi(z^{\prime},t^{\prime})]} \ra.
\eearr
We can write down the contributions from the density fluctuation to the correlation function
as 
$$
\la (1- \frac{1}{8}[\frac{\delta \hat n(z)}{ n_0(z)} -
\frac{\delta \hat n(z^{\prime})}{ n_0(z^{\prime})} ]^2) \ra
 = e^{-\frac{1}{8} \la [\frac{\delta \hat n(z)}{ n_0(z)} -
\frac{\delta \hat n(z^{\prime} )}{ n_0(z^{\prime})}]^2 \ra} = e^{-\frac{F_d}{8}} \nonumber,
$$
where 
\bearr \label{fd}
F_d[z,z^{\prime}] & = &  \la [\frac{\delta \hat n(z)}{ n_0(z)} -
\frac{\delta \hat n(z^{\prime})}{ n_0(z^{\prime})} ]^2 \ra \nonumber \\
& = & \frac{a a_z}{2\sqrt{2}a_0^2} \frac{a_z^5}{Z_T^5} 
\sum_{j=0}^{j_{\rm max}} (2j+1) \sqrt{j(j+1)} coth(\frac{\beta \hbar \omega_j}{2}) 
\nonumber \\ 
& \times & [\frac{P_j(z)}{1-z^2} - \frac{P_j(z^{\prime})}{1-z^{\prime 2}}]^2.
\eearr

Interestingly, the contribution of the density fluctuations given in the above
is similar to the result obtained by Mora and Castin \cite{mora} by discretizing the
low-dimensional space. 

Now we calculate the thermodynamic average of the phase part. 
Following the Ref. \cite{kada}, we denote, 
$ X = -i[ \delta \hat \phi(z,t) - \delta \hat \phi(z^{\prime},t^{\prime})] $, 
then
\be
\la e^X \ra = e^{\sum_{ n = 1} \frac{1}{n} \la X^n \ra_c}.
\ee

The first term and the third term in the exponent is zero since $ \la \hat \alpha_j \ra = 0 $.
The second term is, 
$$
\frac{1}{2} \la X^2 \ra_c = \frac{1}{2} \{ \la X^2 \ra - \la X \ra^2 \} 
= -\frac{1}{2} \la [ \delta \hat \phi(z,t) - \delta \hat \phi(z^{\prime},t^{\prime})]^2 \ra \nonumber. 
$$
The fourth-order term is
$ \frac{1}{24} \la X^4 \ra_c = \frac{1}{24} \{ \la X^4 \ra - 3 \la X^2 \ra^2 \} $. 
We will take the fourth-order term in our following calculations and show that it
contributes significantly to the correlation functions even if we use the current experimental
parameters. Then the single-particle density matrix becomes
$$
D_1(z,z^{\prime}) = \sqrt{n_0(z) n_0(z^{\prime})} e^{-\frac{1}{2} F_2[z,z^{\prime}] 
+ \frac{1}{24} F_4[z,z^{\prime}] - \frac{F_d[z,z^{\prime}]}{8}},
$$
where 
\bearr \label{f2}
F_2[z,z^{\prime}] & = & \frac{a a_z}{a_0^2} \frac{a_z}{Z_T} \sum_{j=0}^{j_{\rm max}} \frac{(2j+1)}{\sqrt{2j(j+1)}}
[P_j(z)-P_j(z^{\prime})]^2 \nonumber \\
& \times & coth(\frac{\beta \hbar \omega_j}{2}),
\eearr
and $ F_4[z,z^{\prime}] =  F_4^{\prime}[z,z^{\prime}] - 3 (F_2[z,z^{\prime}])^2 $,
where
\bearr
F_4^{\prime}[z,z^{\prime}] & = & 3 (\frac{a a_z}{a_0^2} \frac{a_z}{Z_T})^2
 \sum_{j=0}^{j_{\rm max}} \frac{(2j+1)^2}{2j(j+1)}[P_j(z)-P_j(z^{\prime})]^4 \nonumber \\ 
& \times & (1 + 2f_j + 2f_j^2).
\eearr

We calculate all the summation within the phonon regime, $ \hbar \omega_j \leq \mu $ and
the upper cut-off limit, $ j_{\rm max}$, is obtained from the relation, $\mu = \hbar \omega_{j_{\rm max}} $. 
The normalized one-body density matrix or the phase correlation function, $ C_1(\tilde z ) $,
is defined as 
\be
C_1(z, z^{\prime}) = \frac{\la \hat \psi^{\dag} (z) \hat \psi (z^{\prime}) \ra}{\sqrt{n_0(z) n_0(z^{\prime})}} = 
e^{-\frac{1}{2}(F_2 - F_4 / 12 + F_d / 4)}.
\ee
This modified one-body density matrix can also be used for other quasi-condensates,
like quasi-condensate in two-dimensional \cite{pet2} and elongated but three dimensional 
Bose systems \cite{pet3}.
When the system is very close to the zero temperature,
the contributions from $ F_d[z] $ and $F_4[z]$ are negligible, but these two terms 
contribute significantly for a large separations  and at finite temperature ($T > 0.2 T_c $). 
In Fig.6, we show the correlation function with and without those
terms at very high temperature.
\begin{figure}[h]
\epsfxsize 9cm
\centerline {\epsfbox{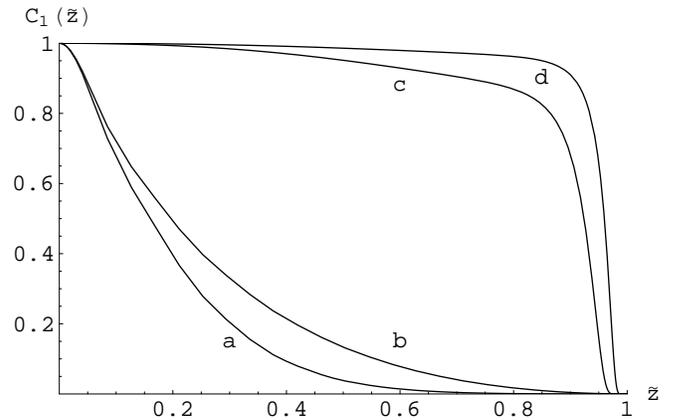}}
\vspace{0.2 cm}
\caption{Plots of the normalized one-body density matrix $ C_1(\tilde z ) $ vs. the
the separation $\tilde z $ (a) at $ T = 0.5 T_c $ and $ F_d = F_4 \neq 0 $, 
(b) $ T = 0.5 T_c $ and $ F_d = F_4 = 0 $, (c) and (d) at $ T = 0.3 T_c$ and $ T = 0.1 T_c $ respectively 
with $a = 2.75 \times 10^{-11} $ and other parameters are fixed.}
\end{figure}
Fig. 6(a) and Fig.6(b) shows that there is a significant difference between the two curves at large separations. 
To describe the correlation function more accurately for a wide range of the
parameters, we must need to include the effect of the 
density fluctuations, fourth-order term of the phase fluctuations and temperature-dependent Thomas-Fermi length
$Z_T$. Our studies on coherence properties improves on 
previous studies in \cite{pet1}.
By taking considerations of density fluctuations and the phase fluctuations upto fourth-order
term, we calculate the phase coherence $ C_1(\tilde z ) $ vs. the separation $ \tilde z $ which 
is shown in the Fig.7.
\begin{figure}[h]
\epsfxsize 9cm
\centerline {\epsfbox{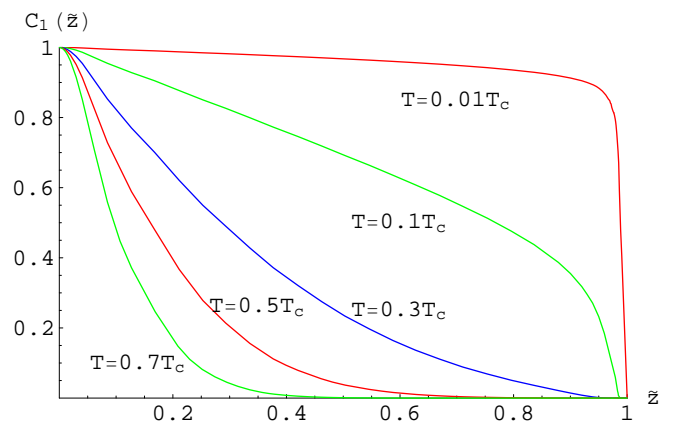}}
\vspace{0.2 cm}
\caption{Plots of the normalized one-body density matrix $ C_1(\tilde z ) $ vs. the
the separation $\tilde z $ for various temperatures.}
\end{figure}

The phase coherence length ($L_p$) of the condensate is 
defined as the distance at which the first
order correlation function is equal to 
$ 1/\sqrt{e} $, i.e., $ C_1(z-z^{\prime} = L_p ) = 1/\sqrt{e} $ which is shown 
in the Fig.8 for various temperature.

\begin{figure}[h]
\epsfxsize 9cm
\centerline {\epsfbox{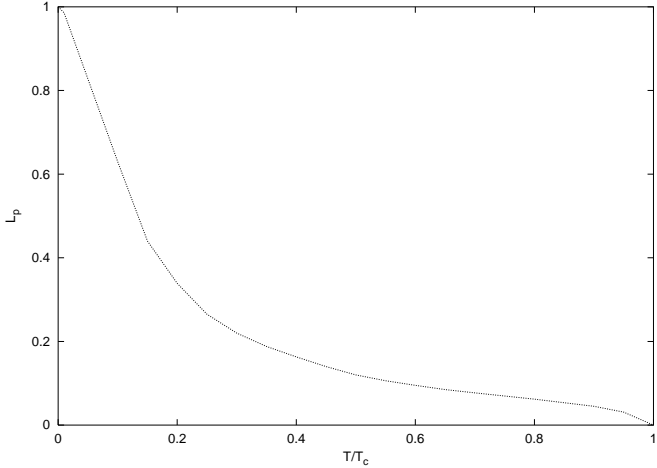}}
\vspace{0.2 cm}
\caption{Plots of the phase coherence length $L_p$ vs. temperature $ T$ in units of $T_c$.}
\end{figure}

The momentum distribution can be obtained from Fourier transformation 
of the one-particle density matrix: 
\be
n(q) = \int_{-Z_{T}}^{Z_{T}} dz \int_{-Z_{T}}^{Z_{T}} dz^{\prime} D_1(z, z^{\prime}) e^{iq(z-z^{\prime})}.
\ee 
Fig.9 shows that the momentum distribution broadens due to the increase of the phase 
and density fluctuations with increasing $T$. There is a long tail in the momentum 
distribution at high temperature, whereas at low temperature, it decreases very fast.
The long-tail in the momentum distribution can be used to identify the presence of a large
phase fluctuations in the bosonic system. 

\begin{figure}[h]
\epsfxsize 9cm
\centerline {\epsfbox{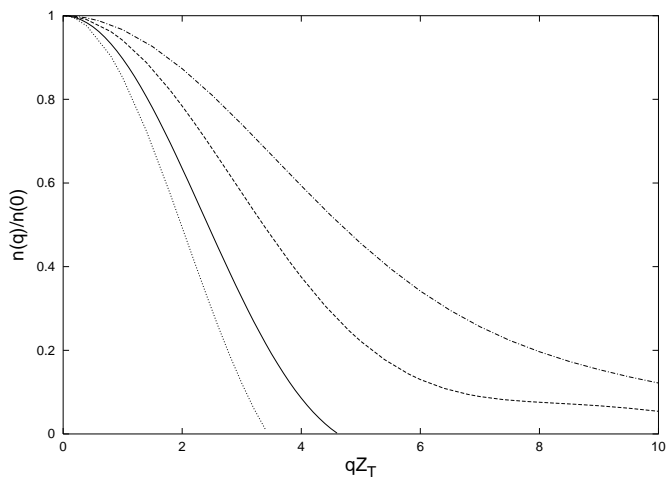}}
\vspace{0.2 cm}
\caption{Plots of the momentum distribution $n(q)/n(0)$ vs. momentum $qZ_T$ for various temperature  
$T=0$ (dotted), $T = 0.1T_c $ (solid), $T = 0.3T_c $ (dashed) and $T = 0.5T_c $ (dot-dashed).}
\end{figure}

The estimated temperature of trapped Bose gases at MIT is $T\sim 0.1 T_c$,
and Fig7. shows that the phase is not coherent overall the system at $T = 0.1 T_c$.
Therefore it forms a quasi-condensate with a large phase fluctuations.
One can see from Eqs. (\ref{fd}) and (\ref{f2}) that
$F_d \sim (\frac{1}{a^2 N_{c0}^5})^{1/3} $, and  
$ F_2 \sim (\frac{a^2}{N_{c0}})^{1/3}$.
It implies that for a given number of particles, strong
interaction suppresses the density fluctuation where as the phase fluctuation is
enhanced. By moderate change of the interaction strength $a$, one can produce a true 
condensate even at temperature $T \sim 0.1-0.2 T_c$. 
In fact, if we use $ a = 2.75 \times 10^{-11} m$  and keeping other parameters fixed as used in other figures, 
we find there is a true condensate where the phase correlation function $C_1(z)$ 
is almost constant (see Fig.6(c) and Fig.6(d)) overall the system even at temperature $T \sim 0.1-0.2 T_c$.
Also, we plot the momentum distribution in Fig.10 and shows that there is no
broadening and long-tail in the momentum distribution even at $ T \sim 0.3 T_c$.

\begin{figure}[h]
\epsfxsize 9cm
\centerline {\epsfbox{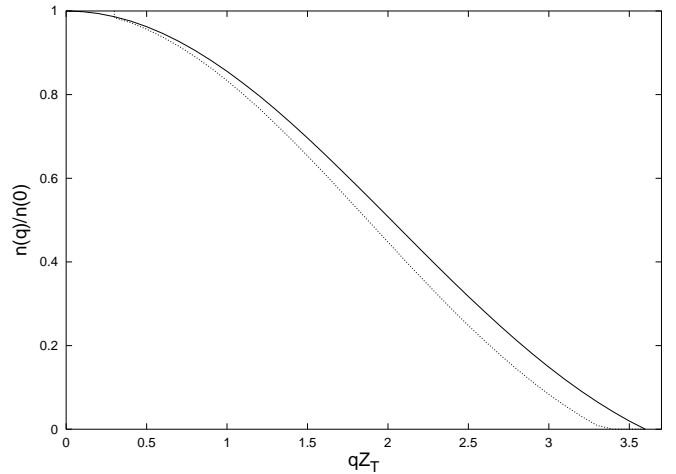}}
\vspace{0.2 cm}
\caption{Plot of the momentum distribution $ n(q)/n(o)$ vs. $ q Z_T$ at $T=0.1T_c$ (dotted) and
at $T=0.3T_c$ (solid) when $ a = 2.75 \times 10^{-11} m$.}
\end{figure}
Note that the new parameters also satisfy the condition for a system to be quasi-one dimensional.

\section{summary}
In this paper, we have presented a detail investigation of the effect of 
density and phase fluctuations on physical observables, like the discrete
spectrums, dynamic structure factor, momentum transfer due to Bragg
scattering and the momentum distribution. 
We have also studied the dynamic structure factor which reflects the discrete nature of the
spectrum and discuss how and when these multipeaks in the dynamic structure factor 
can be observed through the Bragg spectroscopy.
We have calculated the one-body density matrix, correlation length and the momentum distribution
by considering the density fluctuations, phase fluctuations upto quartic term and also the
temperature-dependent Thomas-Fermi length. Our studies improves on the previous studies
where the density fluctuations, higher order phase fluctuations and the temperature-dependent 
Thomas-Fermi length were not considered.  
The coherence properties studied from the density and phase fluctuations reveals that the Bose gas of
MIT experiment do not form a true condensate, but we have shown that a true
condensate can be achieved by a moderate change of the
interaction strength which are accessible due to the Feshbach resonance \cite{fesh}.

\end{multicols}
\end{document}